\documentclass[12pt,nofootinbib]{article}
\pdfoutput=1
\usepackage{soul}
\usepackage{slashed}
\usepackage{amsmath,amssymb,graphicx,multicol} 
\usepackage{epsf}
\usepackage[nosort]{cite}
\usepackage{cancel}
\usepackage{framed}
\usepackage{multirow}
\usepackage{hyperref}
\usepackage{color}

\newcommand{\beq}{\begin{eqnarray}}
\newcommand{\eeq}{\end{eqnarray}}

\newcommand{\centeron}[2]{{\setbox0=\hbox{#1}\setbox1=\hbox{#2}\ifdim

\wd1>\wd0\kern.5\wd1\kern-.5\wd0\fi \copy0

\kern-.5\wd0\kern-.5\wd1\copy1\ifdim\wd0>\wd1
                                       \kern.5\wd0\kern-.5\wd1\fi}}
\newcommand{\ltap}{\>\centeron{\raise.35ex\hbox{$<$}}
                               {\lower.65ex\hbox{$\sim$}}\>}
\newcommand{\gtap}{\>\centeron{\raise.35ex\hbox{$>$}}
                               {\lower.65ex\hbox{$\sim$}}\>}

\newcommand\ZZ{\hbox{\zfont Z\kern-.4emZ}}
\font\zfont = cmss10 %scaled \magstep1

\begin{document}
\begin{titlepage}
\vspace*{-2.5cm}
\begin{flushright}
{\small
CERN-PH-TH/2013-316 \\
KA-TP-03-2014 \\
PSI-PR-13-20
%{\tt hep-ph/yymmnn}
}
\end{flushright}
\vspace*{1.0cm}

\begin{center}
{\Large \bf  
eHDECAY:
an Implementation of\\
the Higgs Effective Lagrangian
 into HDECAY
%Effective Lagrangian for a light Higgs-like scalar
}
\end{center}
\vskip0.5cm

\renewcommand{\thefootnote}{\fnsymbol{footnote}}

\begin{center}
{\large Roberto Contino$^{\,a,b}$\footnote{On leave from Dipartimento di Fisica,  Universit\`a di Roma La Sapienza and INFN, Roma, Italy.}, 
Margherita Ghezzi$^{\, c,d}$, Christophe Grojean$^{\,e}$, \\[0.5cm]
Margarete M\"uhlleitner$^{\, f}$ and Michael Spira$^{\, g}$}
\end{center}

\vskip 20pt

\begin{center}
\centerline{$^{a}$ {\small \it Institut de Th\'eorie des Ph\'enom\`enes Physiques, EPFL, Lausanne, Switzerland}}
\vskip 4pt
\centerline{$^{b}${\small \it Theory Division, Physics Department, CERN, Geneva, Switzerland}}
\vskip 4pt
\centerline{$^{c}$ {\small \it Dipartimento di Fisica,  Universit\`a di Torino and INFN, Torino, Italy}}
\vskip 4pt
\centerline{$^{d}$ {\small \it Dipartimento di Fisica,  Universit\`a di Roma La Sapienza and INFN, Roma, Italy}}
\vskip 4pt
\centerline{$^{e}$ {\small \it ICREA at IFAE, Universitat Aut\`onoma de Barcelona, E-08193 Bellaterra, Spain}}
\vskip 4pt
\centerline{$^{f}${\small \it Institute for Theoretical Physics, Karlsruhe Institute of Technology, Karlsruhe, Germany}}
\vskip 4pt
\centerline{$^{g}${\small \it Paul Scherrer Institut, CH--5232 Villigen PSI, Switzerland}}
\end{center}

\vglue 1.0truecm

\begin{abstract}
\noindent 
We present {\tt eHDECAY}, a modified version of the program {\tt
  HDECAY} which includes the full list of leading bosonic operators of
the Higgs effective Lagrangian with a linear or non-linear realization
of the electroweak symmetry and implements two benchmark composite
Higgs models. 
\end{abstract}
\end{titlepage}

%%%%%%%%%%%%%%%%%%%%%%%%%%%%%%%%%%%%%%%%%%%%%%%%%%%%%%
\section{Introduction}
%%%%%%%%%%%%%%%%%%%%%%%%%%%%%%%%%%%%%%%%%%%%%%%%%%%%%%

In a companion paper~\cite{Contino:2013kra}, we gave a detailed review
of the low-energy effective Lagrangian which describes a light
Higgs-like boson and estimated the deviations induced by the leading
operators  to the Higgs decay rates.
We discussed in particular how the effective Lagrangian can be used
beyond the tree-level by performing a multiple perturbative expansion 
in the SM coupling  parameter $\alpha/\pi$ and in powers of $E/M$,
where $E$ is the energy of the process and $M$ is the New Physics (NP)
scale  at which new massive states appear.
When the Higgs-like boson is part of a weak doublet,  a third
expansion must be performed for $v/f \ll 1$, where $f\equiv M/g_\star$
and $g_\star$ is the typical coupling of the NP sector.

A recent study~\cite{Elias-Miro:2013mua} concluded that, at
tree-level, there are 8 dimension-6 CP-even operators that can be
constrained by Higgs physics only. It is of course essential to have
automatic tools to give accurate predictions of the deviations induced
by these operators to Higgs observables. These operators are all part
of the Strongly Interacting Light Higgs (SILH) Lagrangian~\cite{SILH}
that we will be dealing with (the SILH Lagrangian, Eq.~\ref{eq:silh},
contains 12 operators but 2 combinations of them are severely
constrained by electroweak (EW) precision data and two other combinations are
constrained by the bounds on anomalous triple gauge couplings). 
These operators are also included in Monte Carlo codes recently 
developed~\cite{Artoisenet:2013puc,Alloul:2013naa}.  

The purpose of this note is to present the Fortran code {\tt eHDECAY}, 
which implements the leading operators in the effective Lagrangian and gives 
an extension of the  program {\tt HDECAY}~\cite{hdecay} for the 
automatic calculation of the Higgs decay widths and branching ratios.
The program can be obtained at the URL:\newline 
\centerline{\url{http://www-itp.particle.uni-karlsruhe.de/~maggie/eHDECAY/}.}

The organization of the paper is as follows. In
Section~\ref{sec:efflags} we briefly review the definition of the effective
Lagrangians, with linearly and non-linearly realized electroweak
symmetry breaking (EWSB), 
that have been implemented in the program. This is mainly to set the
notation. For more details and for a discussion of
the physics implications we refer the reader to
Ref.~\cite{Contino:2013kra}. Section~\ref{sec:ehdecay} gives a detailed
discussion of how the partial decay widths have been implemented into
the program {\tt eHDECAY}, 
including higher-order effects in the perturbative expansion.
For issues related to  the perturbative expansion and the inclusion of
higher-order corrections we again refer 
the reader to Ref.~\cite{Contino:2013kra}. In Section~\ref{sec:numerical}
we give numerically approximated results for the Higgs decay rates in
the framework of linearly realized EWSB. Section~\ref{sec:IO} explains 
how to run {\tt eHDECAY} and presents sample input and output
files. We conclude in Section~\ref{sec:concl}.

%%%%%%%%%%%%%%%%%%%%%%%%%%%%%%%%%%%%%%%%%%%%%%%%%%%%%%
\section{Effective Lagrangians for linearly and 
  non-lineary realized EW symmetry}
\label{sec:efflags}
%%%%%%%%%%%%%%%%%%%%%%%%%%%%%%%%%%%%%%%%%%%%%%%%%%%%%%

We assume for simplicity that the Higgs boson is CP-even and that
baryon and lepton numbers are conserved. If  the Higgs is part of a weak doublet,
the leading effects beyond the Standard Model are parametrized by 53
operators with dimension-6~\cite{Buchmuller:1985jz,others,Grzadkowski:2010es} 
(additional 6 operators must be added if the assumption of CP conservation is relaxed), when a single family of quarks and leptons is considered.
In the following we will adopt the so-called SILH basis proposed in
Ref.~\cite{SILH}:
\begin{equation}
\label{eq:effL}
{\cal L} = {\cal L}_{SM}  + \sum_i \bar c_i O_i \equiv {\cal L}_{SM} + \Delta {\cal L}_{SILH}  + \Delta {\cal L}_{F_1} +  \Delta {\cal L}_{F_2}  +
\Delta {\cal L}_{V} + \Delta {\cal L}_{4F}\, ,
%\\
\end{equation}
with~\footnote{In this paper we follow the same notation as in
  Ref.~\cite{Contino:2013kra}. In particular, the expression of the SM
  Lagrangian ${\cal L}_{SM}$ and the convention for the covariant
  derivatives and the gauge field strengths can be found in Appendix~A
  of Ref.~\cite{Contino:2013kra}.
}
\begin{align}
\label{eq:silh}
\begin{split}
\Delta {\cal L}_{SILH} =
\, & \frac{\bar c_H}{2v^2}\, \partial^\mu\!\left( H^\dagger H \right) \partial_\mu \!\left( H^\dagger H \right) 
+ \frac{\bar c_T}{2v^2}\left (H^\dagger {\overleftrightarrow { D^\mu}} H \right) \!\left(   H^\dagger{\overleftrightarrow D}_\mu H\right)  
- \frac{\bar c_6\, \lambda}{v^2}\left( H^\dagger H \right)^3 \\[0.2cm]
& + \left( \left( \frac{\bar c_u}{v^2}\,  y_{u}\, H^\dagger H\,   {\bar q}_L H^c u_R +  \frac{\bar c_d}{v^2}\,  y_{d}\, H^\dagger H\,   {\bar q}_L H d_R 
+ \frac{\bar c_l}{v^2}\,  y_{l}\, H^\dagger H\,   {\bar L}_L H l_R \right)  + {\it h.c.} \right)
\\[0.2cm]
& +\frac{i\bar c_W\, g}{2m_W^2}\left( H^\dagger  \sigma^i \overleftrightarrow {D^\mu} H \right )( D^\nu  W_{\mu \nu})^i
+\frac{i\bar c_B\, g'}{2m_W^2}\left( H^\dagger  \overleftrightarrow {D^\mu} H \right )( \partial^\nu  B_{\mu \nu})   
\\[0.2cm]
&
+\frac{i \bar c_{HW} \, g}{m_W^2}\, (D^\mu H)^\dagger \sigma^i (D^\nu H)W_{\mu \nu}^i
+\frac{i\bar c_{HB}\, g^\prime}{m_W^2}\, (D^\mu H)^\dagger (D^\nu H)B_{\mu \nu} 
\\[0.2cm]
&+\frac{\bar c_\gamma\,  {g'}^2}{m_W^2}\, H^\dagger H B_{\mu\nu}B^{\mu\nu}
   +\frac{\bar c_g \, g_S^2}{m_W^2}\, H^\dagger H G_{\mu\nu}^a G^{a\mu\nu}
\, , 
\end{split}
% \\
\end{align}
%\vspace{0.5cm}
%
\begin{align}
\label{eq:silh2}
\begin{split}
\Delta {\cal L}_{F_1} =
\, & \frac{i \bar c_{Hq}}{v^2}  \left(\bar q_L \gamma^\mu q_L\right)  \big( H^\dagger{\overleftrightarrow D}_\mu H\big)
+ \frac{i \bar c^\prime_{Hq}}{v^2}  \left(\bar q_L \gamma^\mu \sigma^i q_L\right)  \big(H^\dagger\sigma^i {\overleftrightarrow D}_\mu H\big) 
\\[0.2cm]
& + \frac{i \bar c_{Hu}}{v^2}  \left(\bar u_R \gamma^\mu u_R\right)  \big( H^\dagger{\overleftrightarrow D}_\mu H\big)
+ \frac{i \bar c_{Hd}}{v^2}  \left(\bar d_R \gamma^\mu d_R\right)  \big( H^\dagger{\overleftrightarrow D}_\mu H\big) \\[0.2cm]
& +\left(  \frac{i \bar c_{Hud}}{v^2}  \left(\bar u_R \gamma^\mu d_R\right)  \big( H^{c\, \dagger} {\overleftrightarrow D}_\mu H\big) +{\it h.c.} \right)
 \\[0.2cm]
& + \frac{i \bar c_{HL}}{v^2}  \left(\bar L_L \gamma^\mu L_L\right)  \big( H^\dagger{\overleftrightarrow D}_\mu H\big)
+ \frac{i \bar c^\prime_{HL}}{v^2}  \left(\bar L_L \gamma^\mu \sigma^i L_L\right)  \big(H^\dagger\sigma^i {\overleftrightarrow D}_\mu H\big)
 \\[0.2cm]
& + \frac{i \bar c_{Hl}}{v^2}  \left(\bar l_R \gamma^\mu l_R\right)  \big( H^\dagger{\overleftrightarrow D}_\mu H\big) \, ,
\end{split} 
\\[0.7cm]
\label{eq:silh3}
\begin{split}
\Delta {\cal L}_{F_2} =
\, & \frac{\bar c_{uB}\, g'}{m_W^2}\,  y_u \,  {\bar q}_L H^c \sigma^{\mu\nu} u_R \,  B_{\mu\nu}
 + \frac{\bar c_{uW}\, g}{m_W^2}\,  y_u \, {\bar q}_L  \sigma^i H^c \sigma^{\mu\nu} u_R  \, W_{\mu\nu}^i
 + \frac{\bar c_{uG}\, g_S}{m_W^2}\, y_u \,  {\bar q}_L H^c \sigma^{\mu\nu} \lambda^a u_R  \, G_{\mu\nu}^a 
\\[0.2cm]
\, &  +  \frac{\bar c_{dB}\, g'}{m_W^2}\, y_d \, {\bar q}_L H \sigma^{\mu\nu} d_R \, B_{\mu\nu}
+  \frac{\bar c_{dW}\, g}{m_W^2}\,   y_d \, {\bar q}_L \sigma^i H \sigma^{\mu\nu} d_R \, W_{\mu\nu}^i
+  \frac{\bar c_{dG}\, g_S}{m_W^2}\,  y_d \,  {\bar q}_L H \sigma^{\mu\nu}  \lambda^a d_R \, G_{\mu\nu}^a
\\[0.2cm]
\, & + \frac{\bar c_{lB}\, g'}{m_W^2}\,  y_l \,   {\bar L}_L H  \sigma^{\mu\nu}  l_R \, B_{\mu\nu}
+ \frac{\bar c_{lW}\, g}{m_W^2}\,  y_l \,  {\bar L}_L \sigma^i H  \sigma^{\mu\nu}  l_R \, W_{\mu\nu}^i
+ {\it h.c.} \, .
\end{split}
\end{align}
Here $\lambda$ denotes the Higgs quartic coupling which appears in the SM Lagrangian ${\cal L}_{SM}$, and the weak scale 
is defined by 
\begin{equation}
\label{eq:vdef}
v\equiv\frac{1}{(\sqrt{2} G_F)^{1/2}} \simeq 246\,\text{GeV}\, .
\end{equation}
We have defined the Hermitian derivative
\begin{equation}
i H^\dagger {\overleftrightarrow { D^\mu}} H \equiv i H^\dagger
(D^\mu H) - i   (D^\mu H)^\dagger H 
\end{equation}
and $\sigma^{\mu\nu} \equiv i [\gamma^\mu, \gamma^\nu]/2$. 
The Yukawa couplings $y_{u,d,l}$ and the Wilson
coefficients $\bar c_{i}$ are matrices in flavor space, and a
summation over flavor indices has been implicitly assumed.  
In order to avoid large Flavor-Changing
Neutral Currents (FCNC) through the tree-level exchange of the Higgs
boson, we assume that each of the operators $O_{u,d,l}$ is
flavor-aligned with the corresponding mass term. The coefficients
$\bar{c}_{u,d,l}$ are then proportional to the identity matrix in
flavor space. Furthermore, as we assume CP-invariance, they
are taken to be real. A naive estimate of the Wilson coefficients
$\bar c_i$ can be found in Eq.~(2.9) of Ref.~\cite{Contino:2013kra},
following the power counting of Ref.~\cite{SILH}. 

In addition to those listed in Eqs.~(\ref{eq:silh})-(\ref{eq:silh3}),
the effective Lagrangian includes also five extra bosonic operators,
$\Delta {\cal L}_{V}$, as well as 22 four-Fermi baryon-number
conserving operators,~$\Delta {\cal L}_{4F}$. Two of the operators in
Eqs.~(\ref{eq:silh}), (\ref{eq:silh2}) are in fact redundant and can
be eliminated through the equations of motion. A most convenient
choice is that of eliminating two of the three operators  involving
leptons in $\Delta {\cal L}_{F_1}$. 

In the unitary gauge with canonically normalized fields, the SILH
effective Lagrangian $\Delta {\cal L}_{SILH}$ reads: 
\begin{equation} 
\label{eq:chiralL}
\begin{split}
{\cal L} =
& \, \frac{1}{2} \partial_\mu h\ \partial^\mu h - \frac{1}{2} m_h^2 h^2 
- c_3 \, \frac{1}{6} \left(\frac{3 m_h^2}{v}\right) h^3 - \sum_{\psi = u,d,l} m_{\psi^{(i)}} \, \bar\psi^{(i)}\psi^{(i)} \left( 1 + c_\psi \frac{h}{v} + \dots \right)
\\[0.cm]
& + m_W^2\,  W^+_\mu W^{-\, \mu} \left(1 + 2 c_W\, \frac{h}{v} + \dots \right) + \frac{1}{2} m_Z^2\,  Z_\mu Z^\mu \left(1 + 2 c_Z\, \frac{h}{v} + \dots \right) + \dots
\\[0.3cm]
& + \left(  c_{WW}\,   W_{\mu\nu}^+ W^{-\mu\nu}  + \frac{c_{ZZ}}{2} \, Z_{\mu\nu}Z^{\mu\nu} + 
c_{Z\gamma} \, Z_{\mu\nu} \gamma^{\mu\nu}   + \frac{c_{\gamma\gamma}}{2}\, \gamma_{\mu\nu}\gamma^{\mu\nu} + \frac{c_{gg}}{2}\, G_{\mu\nu}^aG^{a\mu\nu} \right) \frac{h}{v}
\\[0.2cm]
& +  \Big( \left(c_{W\partial W}\, W^-_\nu D_\mu W^{+\mu\nu}+ h.c.\right)+c_{Z\partial Z}\,  Z_\nu\partial_\mu Z^{\mu\nu}
+  c_{Z\partial \gamma}\, Z_\nu\partial_\mu\gamma^{\mu\nu} \Big)\,
\frac{h}{v} + \dots
\end{split}
\end{equation}
where we have shown terms with up to three fields and at least one Higgs boson.
The couplings $c_i$ are linear functions of the Wilson coefficients of the effective Lagrangian~(\ref{eq:effL}) and are
reported in Table~\ref{tab:coupvalues}.~\footnote{Notice that the similar Table~1 in Ref.~\cite{Contino:2013kra} contains an
erroneous factor 2 in the dependence of $c_Z$~on~$\bar c_T$.}
%%%%%%%%%%%%%%%%%%%%%%%%%%%%%%%%%%%%%%%%%%%%%%%%%%%%%
\begin{table}
\vspace*{-1cm}
\begin{center}
\renewcommand{\arraystretch}{2.0}
\begin{tabular}{cccc}
\hline
Higgs couplings & $\Delta {\cal L}_{SILH}$ & MCHM4 & MCHM5 \\ \hline
$c_W$ & $1 -\bar c_H/2$ & $\sqrt{1-\xi}$ & $\sqrt{1-\xi}$ \\
$c_Z$ & $1 -\bar c_H/2 - \bar c_T$ & $\sqrt{1-\xi}$ & $\sqrt{1-\xi}$ \\
$c_\psi \ \left(\psi=u,d,l\right)$ & $1- (\bar c_H/2 + \bar c_\psi)$ & $\sqrt{1-\xi}$ & $\displaystyle \frac{1-2\xi}{\sqrt{1-\xi}}$\\
$c_3$ & $1+ \bar c_6 - 3 \bar c_H/2$ & $\sqrt{1-\xi}$ & $\displaystyle \frac{1-2\xi}{\sqrt{1-\xi}}$ \\
$c_{gg}$ & $8\, (\alpha_s/\alpha_2) \,\bar c_{g}$ &  0 & 0 \\
$c_{\gamma\gamma}$ & $8\sin^2\!\theta_W \,\bar c_{\gamma}$& 0 & 0 \\
$c_{Z\gamma}$ & $\displaystyle \left(\bar c_{HB}-\bar c_{HW} - 8\, \bar c_\gamma \sin^2\!\theta_W\right) \tan\theta_W$ & 0 & 0 \\
$c_{WW}$ & $-2\,\bar c_{HW}$ & 0 & 0\\
$c_{ZZ}$ & $-2\left(\bar c_{HW} +  \bar c_{HB} \tan^2\!\theta_W - 4 \bar c_\gamma\tan^2\!\theta_W \sin^2\!\theta_W \right)$ 
& 0 & 0 \\
$c_{W\partial W}$ & $\displaystyle -2(\bar c_W+\bar c_{HW})$ & 0 & 0 \\
$c_{Z\partial Z}$ & $\displaystyle - 2(\bar c_W+ \bar c_{HW}) - 2\left(\bar c_B+\bar c_{HB}\right) \tan^2\!\theta_W$ & 0 & 0 \\
$c_{Z\partial \gamma}$ & $\displaystyle 2\left( \bar c_B+\bar c_{HB} - \bar c_W - \bar c_{HW}\right) \tan\theta_W$ & 0 & 0
\\[0.5cm]
\hline
\end{tabular}
\end{center}
\caption{\small 
The second column reports the values of the Higgs couplings $c_i$
defined in~Eq.~(\ref{eq:chiralL}) in terms of the coefficients $\bar
c_i$ of the effective Lagrangian $\Delta {\cal L}_{SILH}$.  
The last two columns show the predictions of the MCHM4 and MCHM5
models in terms of $\xi = (v/f)^2$, see Ref.~\cite{Contino:2013kra} 
for details. The auxiliary parameter $\alpha_2$ is defined by Eq.~(\ref{eq:alpha2}).
}  
\label{tab:coupvalues}
\end{table}
%%%%%%%%%%%%%%%%%%%%%%%%%%%%%%%%%%%%%%%%%%%%%%%%%%%%%
%
In particular, the following relations hold
\begin{align}
\label{eq:identity1}
c_{WW}  - c_{ZZ} \cos^2\!\theta_W & = c_{Z\gamma} \sin 2\theta_W + c_{\gamma\gamma}  \sin^2\!\theta_W \\[0.4cm]
\label{eq:identity2}
c_{W\partial W}  - c_{Z\partial Z} \cos^2\!\theta_W & = \frac{c_{Z\partial \gamma}}{2} \sin 2\theta_W  \, , 
\end{align}
which are a consequence of the accidental custodial invariance of the
SILH Lagrangian at the level of dimension-6 operators~\cite{Contino:2013kra}.~\footnote{If the assumption of CP conservation
is relaxed, $c_{W\partial W}$ can in general  be complex, while the other bosonic couplings of Eq.~(\ref{eq:chiralL}) are real. In this case
Eq.~(\ref{eq:identity2}) corresponds to two real identities, respectively, on the real and on the imaginary parts, so that 
custodial symmetry implies $\text{Im}(c_{W\partial W}) = 0$.}
Imposing custodial invariance for the Lagrangian~(\ref{eq:silh}), so that $\bar c_T =0$, implies a
third relation that holds for the non-derivative couplings $c_W$ and~$c_Z$:
\begin{equation}
\label{eq:identity3}
c_W =c_Z \, .
\end{equation}
For arbitrary values of the couplings~$c_i$,  Eq.~(\ref{eq:chiralL})
represents the most general effective Lagrangian which can be written at $O(p^4)$  in a derivative expansion by
focusing on cubic terms with at least one Higgs boson 
and making the following two assumptions: \textit{i)} CP is conserved; \textit{ii)} vector fields couple to conserved currents.
Effects which violate the second assumption, in particular, are suppressed by the fermion masses, hence they are small for all the processes
of interest in this work. 
Such description does not require the Higgs boson to be part of an electroweak doublet, and in
this sense Eq.~(\ref{eq:chiralL}) can be considered as a generalization of the SILH Lagrangian $\Delta {\cal L}_{SILH}$. 
It contains 10 couplings involving a single Higgs boson and two gauge fields ($hVV$ couplings, with $V = W,Z,\gamma,g$),
3 linear combinations of which  vanish if custodial symmetry is imposed~\cite{Contino:2013kra}. 
This counting agrees with the complete non-linear Lagrangian at ${\cal O}(p^4)$ recently built in
Refs.\cite{Buchalla:2012qq,Alonso:2012px,Buchalla:2013rka,Brivio:2013pma}. This
general  Lagrangian contains many more operators but it can
be easily checked that only 10 independent operators remain after assuming CP invariance and the conservation
of fermionic currents, and among them 3 break the custodial symmetry. 
If the assumption on conserved currents is relaxed, there are two more independent operators  
at ${\cal O}(p^4)$  that give rise to $hVV$ couplings (they are the operators $P_{9}$ and $P_{10}$ of Ref.~\cite{Brivio:2013pma}, see also the general form factor
description of Ref.~\cite{Isidori:2013cla}). These two additional couplings can only be obtained from
dimension-8 operators when the Higgs boson is part of an EW  doublet.
In the non-linear realization of the EW symmetry,
all Higgs couplings  are truly independent of other parameters that do
not involve the Higgs boson, like EW oblique parameters or anomalous triple
gauge couplings. In a linear realization, on the other hand, only 4 $hVV$
couplings are independent of the other EW measurements~\cite{Elias-Miro:2013mua}. 
In custodial invariant scenarios, it is thus not possible to tell whether the Higgs is part of an EW doublet by  focusing only on $hVV$ couplings,
since their number is the same in both the linear and non-linear descriptions under our assumptions (CP  and current conservation).
The decorrelation between the $hVV$ couplings and the other EW data might instead be a
way to disprove the doublet nature of the Higgs boson~\cite{Brivio:2013pma}.  

The code {\tt eHDECAY} retains only the couplings induced by the operators of $\Delta {\cal L}_{SILH}$ since the effects of the other operators with fermions are either severely constrained by non-Higgs physics or, like the top dipoles, are irrelevant for the Higgs total decay rates (they could modify in a sensible way the differential decay rates but such an analysis is beyond the scope of the present work). The CP-odd operators are not considered either since they do not interfere with the inclusive SM amplitudes and thus modify the decay rates at a subleading order in the perturbative expansion considered in this paper.

%%%%%%%%%%%%%%%%%%%%%%%%%%%%%%%%%%%%%%%%%%%%%%%%%%%%%%
\section{Implementation of the Higgs effective Lagrangian into 
{\tt eHDECAY}  \label{sec:ehdecay}}
%%%%%%%%%%%%%%%%%%%%%%%%%%%%%%%%%%%%%%%%%%%%%%%%%%%%%%
The program {\tt HDECAY}~\cite{hdecay} was originally written for the
automatic computation of the Higgs partial decay widths and branching
ratios in the SM and in its Minimal Supersymmetric extension (MSSM). It
includes the possibility of specifying modified couplings for up-type
quarks, down-type quarks, 
leptons and vector bosons in the parametrization of Eq.~(\ref{eq:chiralL}),  
as well as of including the effective couplings $c_{gg}$,
$c_{\gamma\gamma}$ and $c_{Z\gamma}$. 
We present here a modified version of the program, labelled {\tt
  eHDECAY}. It is available  at the following URL:  \\
\centerline{\url{http://www-itp.particle.uni-karlsruhe.de/~maggie/eHDECAY/}.}\\
In addition to the features already present in {\tt HDECAY}, 
the new program includes the effective 
couplings $c_{WW}$, $c_{ZZ}$, $c_{W\partial W}$ and $c_{Z\partial
  Z}$, and thus fully implements the non-linear
Lagrangian~(\ref{eq:chiralL}).~\footnote{Notice that the operator
  proportional to $c_{Z\partial  \gamma}$ does not affect the decay
  $h\to Z\gamma$ as long as the photon is on-shell.}  
In fact, similarly to  {\tt HDECAY}~5.10, it also includes the
possibility of choosing different couplings of the Higgs boson to each
of the up and down quark flavors and lepton flavors. In this sense the
program assumes neither custodial symmetry nor flavor alignment.
As explained in the text, Eq.~(\ref{eq:chiralL}) describes a generic
CP-even scalar~$h$ at $O(p^4)$ in the derivative expansion. 
If $h$ forms an $SU(2)_L$ doublet together with the longitudinal
polarizations of the $W$ and the $Z$, the Lagrangian can be expanded 
as in Eq.~(\ref{eq:silh}) for $(v/f)\ll 1$; in this case the values
of the Higgs couplings $c_i$ are given in the second column of
Table~\ref{tab:coupvalues}.  The program  {\tt eHDECAY} provides 
an option in its input file where the user can switch from the
non-linear parametrization of Eq.~(\ref{eq:chiralL}) to that of the
SILH Lagrangian Eq.~(\ref{eq:silh}). The user can also choose to set
the values of the Higgs couplings to those predicted at 
leading order in an expansion in powers of weak couplings
% tree-level 
in the benchmark composite Higgs models MCHM4~\cite{Agashe:2004rs}  and
MCHM5~\cite{Contino:2006qr}, see the last two columns of
Table~\ref{tab:coupvalues}. 
 
Similarly to the original version of {\tt HDECAY}, all the relevant
QCD corrections are included. They generally factorize with respect to
the expansion in the number of fields and derivatives of the effective
Lagrangian, and can thus be straightforwardly included by making use
of the existing SM computations. The inclusion of the electroweak
corrections is less straightforward and can currently be done in a
consistent way only in the framework of the Lagrangian~(\ref{eq:silh})
and up to higher orders in $(v/f)$. Going beyond such approximations
would require  dedicated computations which at the moment are not
available in the literature. In {\tt eHDECAY} the user has the option
to include the one-loop EW corrections to a given decay rate only if
the parametrization of Eq.~(\ref{eq:silh}) has been chosen. The same
EW scheme as used by  {\tt HDECAY}, with $G_F$, $m_W$ and $m_Z$ taken
as input parameters, is also adopted in  {\tt eHDECAY}.  The sine of
the Weinberg angle is defined as 
\begin{equation}
\sin^2\!\theta_W = 1 - \frac{m_W^2}{m_Z^2}\, ,
\end{equation}
following the conventional on-shell scheme~\cite{Sirlin:1980nh}. Derived
quantities in this scheme are also the electromagnetic coupling and
the weak coupling. To describe the latter we have conveniently
defined the parameter  
\begin{equation}
\label{eq:alpha2}
\alpha_2 \equiv \frac{\sqrt{2} G_F m_W^2}{\pi}\, .
\end{equation}
The formulas implemented in the program are thus written
in terms of only the input parameters or their derived quantities
$\sin\theta_W$ and $\alpha_2$. 
The only exception to this rule is given by the decay rates
$\Gamma(h\to \gamma\gamma)$ and $\Gamma(h\to Z\gamma)$, where 
we use the experimental value of the electromagnetic coupling in the
Thomson limit, $\alpha_{em}(q^2 =0)$, in order to avoid large
logarithms for on-shell photons.  

Below a detailed discussion follows of how the New Physics
corrections are incorporated for each of the Higgs decay modes. 
We report explicitly the formulas implemented in the code and their
level of approximation in the perturbative expansion of the effective
Lagrangian. In all the following expressions, as explained in the
text, the coefficients of the dimension-6 operators of the SILH
Lagrangian~(\ref{eq:silh}) and those of the derivative operators of
Eq.~(\ref{eq:chiralL}) must be identified with their values at the
relevant low-energy scale $\mu = m_h$. 

%%%%%%%%%%%%%%%%%%%%%%%%%%%%%%%%%%%%%%%%%%%%%%%%%%%%%%%%
\subsection{Decays into quarks and leptons}

Upon adopting the effective description of the non-linear Lagrangian~(\ref{eq:chiralL}) and working at leading order in the derivative
expansion, the Higgs boson partial decay width into a pair of fermions
is obtained by rescaling the tree-level SM value $\Gamma^{SM}_0 (\psi \bar\psi)$ 
by a factor $c_\psi^2$. The QCD corrections to the decay widths into
quarks which are currently available for the SM case include fully
massive next-to-leading order (NLO) corrections near threshold~\cite{nlopsipsi} and massless $\mathcal{O}(\alpha_s^4)$ corrections
far above threshold~\cite{nnlopsipsi,Chetyrkin:1995pd,n3lopsipsi,n4lopsipsi}. Also, large
logarithms can be resummed through the running  of the quark masses
and  of the strong coupling constant.
They are evaluated at the scale given by the Higgs mass. The
transition from the threshold region involving mass effects to the
renormalization-group-improved large-Higgs mass regime is provided by
a smooth linear interpolation.
All these QCD corrections factorize with respect to the tree-level
amplitude and can therefore be incorporated as done in {\tt HDECAY}
for the SM case. The decay rate can be written as follows:
\begin{equation}
\Gamma(\bar\psi\psi)\big|_{NL} = c_\psi^2 \,
\Gamma^{SM}_0(\bar\psi\psi)   \left[ 1 + \delta_\psi \, \kappa^{QCD}
\right] \, , \label{eq:gampsipsi}
\end{equation}
where $ \Gamma^{SM}_0$ denotes the leading-order decay width,
$\delta_\psi =1 (0)$ for $\psi=$ quark (lepton) and $\kappa^{QCD}$
encodes the QCD corrections.\footnote{There is one caveat, however. In
  the case of decays into strange, charm or bottom quarks there are
  two-loop diagrams which involve loops of top quarks coupling to the
  Higgs boson. They need a rescaling different from $c_\psi^2$. It has
  been correctly taken into account by the appropriate modification
  factor $c_\psi c_t$ ($\psi=c,s,b$).}
This is the formula implemented by  {\tt eHDECAY} in the case of the
non-linear Lagrangian~(\ref{eq:chiralL}). It is valid up to
corrections of $O(m_h^2/M^2)$ in the derivative expansion and of
$O(\alpha_2/4\pi)$ from EW loops. These latter corrections are
available in the SM but contrary to the QCD ones do not
factorize. Their inclusion in the case of generic Higgs couplings thus
requires a dedicated calculation, which is not available at
present. The two benchmark composite Higgs models MCHM4 and MCHM5
provide a resummation of 
higher-order terms in
$\xi=v^2/f^2$.
Contrary to
the SILH Lagrangian which is to be seen as an expansion in $\xi$, in
these two models rather large coupling deviations can in principle be
possible (eventually they are precluded due to the constraints from
electroweak precision measurements). We therefore apply the formula
Eq.~(\ref{eq:gampsipsi}) also for the MCHM4 and MCHM5, with $c_\psi$
given by the corresponding coupling values in columns 3 and 4 of
Table~\ref{tab:coupvalues}. 

In case of the SILH parametrization, where the deviations of the Higgs
couplings from their SM  values are assumed to be of $O(v^2/f^2)$ and
small, the decay rate can be written as 
\begin{equation}
\label{eq:fermiondecay}
\Gamma(\bar\psi\psi)\big|_{SILH} = \Gamma^{SM}_0(\bar\psi\psi)  \left[
  1 - \bar{c}_H - 2 \bar{c}_\psi +
  \frac{2}{|A_0^{SM}|^2} \mbox{Re} \left(A_0^{*SM} A^{SM}_{1,ew} \right) \right]
\left[ 1 + \delta_\psi \, \kappa^{QCD} \right] \, ,
\end{equation}
where $A_0^{SM}$, $A^{SM}_{1,ew}$ are, respectively, the tree-level
and EW one-loop~\cite{nloewpsipsi} amplitudes of the~SM. In this case
the one-loop EW corrections can be easily included if one neglects
terms of $O[(\alpha_2/4\pi) (v/f)^2]$~\footnote{\label{footnote:EW}As
  pointed out in footnote 21 of Ref.~\cite{Contino:2013kra}, in the
  strict sense this equation is valid for the genuine EW corrections
  only, while for simplicity we include the (IR-divergent) virtual QED
  corrections to the SM amplitude in the same way. The corresponding
  real photon radiation contributions to the decay rates are treated
  in terms of a {\it linear} novel contribution to the Higgs coupling
  for the squared amplitude in order to obtain an infrared finite
  result. Pure QED corrections factorize as QCD corrections in general
  so that their amplitudes scale with the modified Higgs
  couplings. However, they cannot be separated from the genuine EW
  corrections in a simple way.}.  
In particular, mixed contributions up to $O[ (\alpha_2/4\pi)
(\alpha_{s}/4\pi)^4]$ have been included by assuming that the
electroweak and QCD corrections factorize, as the non-factorizable
contributions are small. From the viewpoint of the expansion in
inverse powers of the NP scale, the formula~(\ref{eq:fermiondecay})
includes corrections of order $O(v^2/f^2)$. It neglects terms of
$O(v^4/f^4)$, $O[(\alpha_2/4\pi) (v/f)^2]$, $O[(\alpha_2/4\pi)^2]$. 
 
%%%%%%%%%%%%%%%%%%%%%%%%%%%%%%%%%%%%%%%%%%%%%%%%%%%%%%%%
\subsection{Decay into gluons}
%%%%%%%%%%%%%%%%%%%%%%%%%%%%%%%%%%%%%%%%%%%%%%%%%%%%%%%%
Upon selecting the Lagrangian~(\ref{eq:chiralL}), 
the rate into two gluons is computed in {\tt eHDECAY} by means of the
following formula: 
\begin{equation}
\label{eq:hggqcd}
\begin{split}
\Gamma(gg)\big|_{NL}  = \frac{G_F\alpha_s^2 m_h^3}{4\sqrt{2}\pi^3} \Bigg[ 
& \bigg| \sum_{q=t,b,c} \frac{c_q}{3} \, A_{1/2}\left(\tau_q\right) \bigg|^2  c_{eff}^2 \, \kappa_{soft}  \\
& + 2\, \text{Re}\!\left( \sum_{q=t,b,c} \frac{c_q}{3} \, A^*_{1/2}\left(\tau_q\right) \frac{2 \pi c_{gg}}{\alpha_s} \right) c_{eff} \, \kappa_{soft}  
    + \left|\frac{2\pi c_{gg}}{\alpha_s}\right|^2  \kappa_{soft}  \\
& + \frac{1}{9} \sum_{q,q'=t,b} c_q\, A^*_{1/2}\left(\tau_q\right) c_{q'}\, A_{1/2}\left(\tau_{q'}\right) \kappa^{NLO} (\tau_q,\tau_{q'}) \Bigg] \, , 
\end{split}
\end{equation}
where $\tau_q=4 m_q^2/m_h^2$ and the loop function, normalized to $A_{1/2}(\infty)=1$,  is defined as
\begin{equation}
 A_{1/2}\left(\tau\right)= \frac{3}{2} \tau\left[1+\left(1-\tau\right)f\left(\tau\right)\right],
\label{eq:A1/2}
\end{equation}
with
\begin{equation}
 f\left(\tau\right)=\left\lbrace \begin{array}{ll}
\displaystyle \arcsin^2\frac{1}{\sqrt{\tau}} & \quad \tau\geq1 \\[0.5cm]
\displaystyle -\frac{1}{4}\left[\ln\frac{1+\sqrt{1-\tau}}{1-\sqrt{1-\tau}}-i\pi\right]^2 & \quad \tau<1\, . \end{array} \right. 
\label{eq:ftau}
\end{equation}
The first term corresponds to the one-loop contribution from the top,
bottom and charm quarks, whose couplings to the Higgs boson are
modified with respect to their SM values. 
In order to minimize the effects from higher-order QCD
  corrections, we use the pole masses for the top, bottom and charm
  quarks, $m_t =172.5$\,GeV, $m_b = 4.75$\,GeV and $m_c = 1.42$\,GeV.
The second and third
terms encode the effect of the derivative interaction between the
Higgs boson and two gluons generated by New Physics. Naively $c_{gg}
\approx (\alpha_s/4\pi) (g_*^2 v^2/M^2)$, so that the correction from
the effective interaction can be as important as the one from the top
quark if $(g_*^2 v^2/M^2) \approx 1$.  No expansion is thus possible
in $c_{gg}$ in the general case. 

The QCD corrections have been included up to N$^3$LO in
Eq.~(\ref{eq:hggqcd}) in the limit of large loop-particle masses,
similarly to what is done in  {\tt HDECAY} for the SM. In this limit
the effect of soft radiation factorizes and is encoded by the
coefficient $\kappa_{soft}$. The coefficient $c_{eff}$, instead,
takes into account the correction from the exchange of hard gluons and
quarks with virtuality $q^2 \gg m^2_t$. More in detail, for $m_h \ll 2
m_t$, one can integrate out the top quark and obtain the following
five-flavour effective Lagrangian 
\begin{equation}
{\cal L}_{\scriptsize \mbox{eff}} = -2^{1/4} G_F^{1/2} C_1 \,
G^0_{a\mu\nu} G_a^{0\mu\nu} h \;,
\end{equation}
where bare fields are labeled by the superscript $0$. The renormalized
coefficient function $C_1$ encodes the dependence on the top quark
mass $m_t$. The coefficients $\kappa_{soft}$ and $c_{eff}$ are thus defined as
\begin{equation}
\begin{split}
\kappa_{soft}   & =   \frac{\pi}{2 m_h^4} \, \text{Im}\,\Pi^{GG} (q^2
= m_h^2) \\[0.2cm] 
c_{eff} & = -\frac{12 \pi \, C_1}{\alpha_s^{(5)} (m_h)} \, ,
\end{split}
\end{equation}
where $\Pi^{GG} (q^2)$ is the vacuum polarization induced by the gluon operator.
The N$^3$LO expression of the coefficient function 
$C_1$~\cite{Chetyrkin:1997un,Kramer:1996iq,Schroder:2005hy,Chetyrkin:2005ia}
in the on-shell scheme and that of $\text{Im}\,\Pi^{GG}$ can be found
in Ref.~\cite{Baikov:2006ch}.
At NLO the expressions for $\kappa_{soft}$ and $c_{eff}$ take the
well-known form~\cite{nloggqcd} 
\begin{equation}
\kappa_{soft}^{NLO} = 1 + \frac{\alpha_s}{\pi} \left(\frac{73}{4} - \frac{7}{6} N_F \right)\, , \qquad
c_{eff}^{NLO} = 1 + \frac{\alpha_s}{\pi} \frac{11}{4} \, ,
\end{equation}
where here $\alpha_s$ is evaluated at the scale $m_h$ and computed for
$N_F=5$ active flavours. In {\tt eHDECAY} it is consistently computed up to
N$^3$LO. 
The last line in Eq.~(\ref{eq:hggqcd}) contains the additional mass
effects at NLO QCD~\cite{Spira:1995rr} in the top and bottom loops, encoded in
$\kappa^{NLO} (\tau_q,\tau_{q'})$, which have been
explicitly implemented in {\tt HDECAY} and taken over in {\tt
  eHDECAY}. While the mass effects for the top quark loops play only a
minor role, below the percent level, for the bottom loop contribution
the mass effects for a 125\,GeV Higgs boson amount to about 8\%
relative to the approximate NLO result.
Hence, formula (\ref{eq:hggqcd}) includes the QCD corrections at N$^3$LO (i.e. at $O(\alpha_s^5)$ in the decay rate),
and neglects next-to-leading order terms  in the derivative expansion (i.e. terms further suppressed by $O(m_h^2/M^2)$).
The decay width within the MCHM4 and MCHM5 is calculated with the same
formula~(\ref{eq:hggqcd}) by replacing $c_q$ with the values in column 3 and 4
of Table~\ref{tab:coupvalues} and $c_{gg} \equiv 0$. 

When the SILH Lagrangian~(\ref{eq:silh}) is selected, on the other
hand,  {\tt eHDECAY}  computes the decay rate into gluons by means of
the following approximate formula: 
\begin{equation}
\begin{split}
\Gamma (gg)\big|_{SILH} = \frac{G_F\alpha_s^2 m_h^3}{4\sqrt{2}\pi^3} \Bigg[ 
& \frac{1}{9} \sum_{q,q'=t,b,c}  (1-\bar{c}_H - \bar{c}_q- \bar{c}_{q'})  \, A^*_{1/2}\left(\tau_{q'}\right) A_{1/2}\left(\tau_q\right)  \,  c_{eff}^2  \, \kappa_{soft} \\[0.1cm]
& + 2\, \text{Re}\!\left( \sum_{q=t,b,c} \frac{1}{3} \, A^*_{1/2}\left(\tau_q\right) \frac{16\pi \, \bar{c}_{g}}{\alpha_2} \right) \, c_{eff} \, \kappa_{soft} \\[0.1cm]
& +  \bigg|\sum_{q=t,b,c} \frac{1}{3} \, A_{1/2}\left(\tau_q\right) \bigg|^2 \, c_{eff}^2 \,  \kappa_{ew} \, \kappa_{soft} \\[0.1cm]
& +  \frac{1}{9} \sum_{q,q'=t,b} (1-\bar{c}_H - \bar{c}_q- \bar{c}_{q'}) A^*_{1/2}\left(\tau_q\right) 
        A_{1/2}\left(\tau_{q'}\right) \kappa^{NLO} (\tau_q,\tau_{q'}) \Bigg] \, .
\end{split}
\end{equation}
The last line contains the mass effects at NLO QCD for the top and
bottom quark loops. The NLO electroweak corrections~\cite{nloewgg1,nloewgg2} are included through the coefficient
$\kappa_{ew}$ and by neglecting terms of 
$O[(\alpha_2/4\pi) (v^2/f^2)]$. The above formula thus includes the
leading $O(v^2/f^2)$ corrections, as well as mixed
$O[(\alpha_s/4\pi)^5 (\alpha_2/4\pi)]$ ones. Indeed, we assume
factorization of the QCD and EW corrections. Since  QCD corrections
are dominated by soft gluon radiation, in which QCD and EW effects
completely factorize, this is a good approximation\footnote{Bottom
  loops contribute $O(10\%)$ to the SM decay rate and are well
  approximated by an effective coupling at the 10\%-level thus leading
  to negligible non-factorizing contributions at the percent level.}.
It neglects terms of $O[(\alpha_2/4\pi)^2]$ and $O(v^4/f^4)$.

%%%%%%%%%%%%%%%%%%%%%%%%%%%%%%%%%%%%%%%%%%%%%%%%%%%%%%%%
\subsection{Decay into photons}
%%%%%%%%%%%%%%%%%%%%%%%%%%%%%%%%%%%%%%%%%%%%%%%%%%%%%%%%
In the SM the decay of the Higgs boson into a pair of photons is
mediated by $W$ and heavy fermion loops. According to the chiral Lagrangian~(\ref{eq:chiralL}), these two
contributions to the total amplitude are rescaled, respectively, by
the parameters $c_W$ and $c_\psi$. Similarly to  $h\rightarrow gg$, 
the contact interaction proportional to $c_{\gamma\gamma}$ can also
contribute significantly. With 
$c_{\gamma\gamma} \approx (\alpha_{em}/4\pi) (g_*^2 v^2/M^2)$, the
contribution due to the effective interaction becomes comparable to
the loop induced contributions if $(g_*^2 v^2/M^2) \approx 1$. 
The partial width for a Higgs boson decaying into two photons
implemented in  {\tt eHDECAY} in the framework of the non-linear
Lagrangian is thus given by  
\begin{equation}
\label{eq:gamdecay}
\begin{split}
\Gamma (\gamma\gamma)\big|_{NL} =
\frac{G_F\alpha_{em}^2m_h^3}{128\sqrt{2}\pi^3}\bigg| 
  & \sum_{q=t,b,c} \frac{4}{3} c_q \, 3 Q_q^2 \, A_{1/2}^{NLO} \left(\tau_q \right)
    + \frac{4}{3} c_\tau Q_\tau^2A_{1/2} \left(\tau_\tau \right)  \\
 & + c_W A_1\left(\tau_W\right)+\frac{4\pi}{\alpha_{em}} c_{\gamma\gamma}\bigg|^2 \, ,
\end{split}
\end{equation}
which is approximate at leading order in the derivative expansion,
\text{i.e.} it neglects terms further suppressed by $O(m_h^2/M^2)$. By $Q_{q,\tau}$ we
denote the electric charge of the quarks and the $\tau$ lepton,
respectively. Note that $\alpha_{em}$ is the electromagnetic coupling
in the Thomson limit, in order to avoid large logarithms for on-shell
photons. We have defined  $\tau_i=4m_i^2/m_h^2$ ($i=q,\tau,W$) and
the form factor 
\begin{equation}
 A_1\left(\tau\right)=-\left[ 2+3\tau+3\tau\left(2-\tau\right)f\left(\tau\right)\right]
\end{equation}
normalized to $A_1(\infty)=-7$.
The top, bottom and charm quark loops receive NLO QCD
corrections, while the effective contact interaction does
not. The NLO QCD corrected quark form factor is denoted in
Eq.~(\ref{eq:gamdecay}) by 
\begin{equation}
A_{1/2}^{NLO} (\tau_q) = A_{1/2} (\tau_q) (1+ \kappa_{QCD}) \, , 
\end{equation}
where $\kappa_{QCD}$ encodes the $O(\alpha_s/4\pi)$  QCD
corrections~\cite{Spira:1995rr,Djouadi:1993ji,nlogagaqcd}
and $A_{1/2}(\tau)$ is given in Eq.~(\ref{eq:A1/2}). 
In the MCHM4 and MCHM5 we use the same formula 
for the decay width with $c_q$ and $c_V$ replaced appropriately and
$c_{\gamma\gamma} \equiv 0$.

In order to improve the perturbative behaviour of the QCD-corrected 
quark loop contributions, they are expressed in terms of
the running quark masses $m_Q(\mu_Q^2)$~\cite{Spira:1995rr,Djouadi:1993ji}. 
These are  related to the pole masses $M_Q$ through
\beq
m_Q (\mu_Q^2) = M_Q \left[ \frac{\alpha_s (\mu_Q^2)}{\alpha_s
    (M_Q^2)}\right]^{12/(33-2N_F)} \left( 1 + {\cal O}(\alpha_s^2)\right)
\eeq
at the mass renormalization point $\mu_Q$ with $N_F=5$ active
flavours. Their scale is identified with $\mu_Q=M_H/2$.  This ensures a
proper definition of the $Q\bar{Q}$ thresholds $M_H=2 M_Q$ without
artificial displacements due to finite shifts between the pole and the
running quark masses, as is the case for the running
$\overline{\mbox{MS}}$ masses. Note, that the same running quark mass $m_Q
(\mu_Q^2)$, at the renormalization scale $\mu_Q=M_H/2$, enters in the lowest
order amplitude $A_{1/2}^{LO}$, which is used in the SILH 
parametrization hereafter.~\footnote{For a Higgs mass value of
  $M_H=125$\,GeV the running top, bottom and charm quark masses are
  given by $m_t = 188.03$\,GeV, $m_b = 3.44$\,GeV and $m_c
  =0.76$\,GeV. They differ from the running $\overline{\mbox{MS}}$
  masses.} 

In the case of the SILH parametrization, the EW
corrections have been incorporated as well. It is useful to define the SM amplitude
at  leading order (LO) and NLO QCD level as
\begin{equation}
A^{SM}_X (\gamma\gamma) = 
\sum_{q=t,b,c} \frac{4}{3} \, 3 Q_q^2 \, A_{1/2}^{X} \left(\tau_q \right)
+ \frac{4}{3} Q_\tau^2 \, A_{1/2} \left(\tau_\tau \right)
+A_1\left(\tau_W\right) \, , \qquad X=LO,NLO \, , 
\end{equation}
and the deviation from the SM amplitude as 
\begin{equation}
\begin{split}
\Delta A (\gamma\gamma) =& - \sum_{q=t,b,c}  \frac{4}{3}
\left(\frac{\bar{c}_H}{2}+\bar{c}_q \right) 3 Q_q^2 \, A_{1/2}^{NLO} \left(\tau_q \right)
- \left(\frac{\bar{c}_H}{2}+\bar{c}_\tau \right) \frac{4}{3} Q_\tau^2
\, A_{1/2} \left(\tau_\tau \right)  \\[0.1cm]
& - \left(\frac{\bar{c}_H}{2} -2 \bar{c}_W \right)
A_1\left(\tau_W\right) \, . 
\end{split}
\end{equation}
The decay width implemented in  {\tt eHDECAY} in the SILH case  is
thus  the following 
\begin{equation}
\label{eq:htogagaSILH}
\begin{split}
\Gamma (\gamma\gamma)\big|_{SILH} =
\frac{G_F\alpha_{em}^2m_h^3}{128\sqrt{2}\pi^3}  \, \Bigg\{ 
& |A^{SM}_{NLO} (\gamma\gamma)|^2 + 2\, \text{Re} \Big( A_{LO}^{SM *} (\gamma\gamma) \, A_{ew}^{SM} (\gamma\gamma) \Big) \\
&  + 2\, \text{Re}\!\left[ A^{SM*}_{NLO} (\gamma\gamma)\left(  \Delta A(\gamma\gamma) 
     + \frac{32\pi \sin^2\!\theta_W \,\bar{c}_{\gamma}}{\alpha_{em}} \right) \right]  \Bigg\} \, ,
\end{split}
\end{equation}
where $A_{ew}^{SM} (\gamma\gamma)$ denotes the SM amplitude which comprises the NLO electroweak 
corrections~\cite{nloewgg1,Degrassi:2005mc}.
Equation~(\ref{eq:htogagaSILH}) includes the leading $O(v^2/f^2)$ and $O(m_h^2/M^2)$ corrections, while it neglects terms of order $O(v^4/f^4)$.  
The electroweak corrections are
implemented up to NLO, neglecting corrections of $O [(\alpha_2/4\pi)(v^2/f^2)]$
and of $O[(\alpha_2/4\pi)^2]$. Finally, the QCD corrections are
included up to NLO, and mixed terms of $O[(\alpha_2/4\pi) (\alpha_s/4\pi)]$ are neglected.

%%%%%%%%%%%%%%%%%%%%%%%%%%%%%%%%%%%%%%%%%%%%%%%%%%%%%%%%
\subsection{Decay into $Z\gamma$}
%%%%%%%%%%%%%%%%%%%%%%%%%%%%%%%%%%%%%%%%%%%%%%%%%%%%%%%%
In the SM the Higgs boson decay into a $Z$ boson and a photon is mediated by 
$W$ boson and heavy fermion loops. Adopting the parametrization
of the non-linear Lagrangian,  the correction from the effective
interaction due to the coupling $c_{Z\gamma}$ has to be considered, too, and
it can become as important as the loop contributions for $(g_*^2
v^2/M^2) \approx 1$.
The decay width is therefore given by (here also $\alpha_{em} \equiv
\alpha_{em} (0)$):
\begin{equation}
\label{eq:Zgammachiral}
\begin{split}
\Gamma (Z\gamma)\big|_{NL}  = 
 & \, \frac{G_F^2\alpha_{em}m_W^2m_h^3}{64\pi^4}  \left(1-\frac{m_Z^2}{m_h^2}\right)^3  \\ 
 & \times  \left|\sum_\psi\frac{c_\psi N_c Q_\psi\hat v_\psi}{\cos\theta_W}\, A_{1/2}^{Z\gamma}\left(\tau_\psi,\lambda_\psi\right)
    +c_W \,
    A_1^{Z\gamma}\left(\tau_W,\lambda_W\right)-\frac{4\pi}{\sqrt{\alpha_{em}
      \alpha_2}} \,  c_{Z\gamma} \right|^2 \, ,
\end{split}
\end{equation}
with $\tau_i=4m_i^2/m_h^2$, $\lambda_i=4m_i^2/m_Z^2$ and $\hat
v_\psi=2I_\psi^3-4Q_\psi\sin^2\theta_W$ ($\psi=t,b,c,\tau$) in terms of
the third component of the weak isospin $I^3_\psi$ and the electric
charge $Q_\psi$. The form factors are defined by\cite{Cahn:1978nz}
\begin{equation}
\begin{split}
A_{1/2}^{Z\gamma}\left(\tau,\lambda\right)=& \left[I_1\left(\tau,\lambda\right)-I_2\left(\tau,\lambda\right)\right]\,, \\[0.2cm]
A_1^{Z\gamma}\left(\tau,\lambda\right)=& \cos\theta_W
\Big\lbrace 4\big(3-\tan^2\theta_W\big)I_2\big(\tau,\lambda\big) \\
&  \hspace{1.4cm} + \Big[\Big(1+\frac{2}{\tau}\Big) \tan^2 \theta_W - \Big(5+\frac{2}{\tau}\Big) \Big]I_1\big(\tau,\lambda\big)
\Big\rbrace \,. 
\end{split}
\end{equation}
The functions $I_1$ and $I_2$ can be cast into the form
\begin{equation}
\begin{split}
&I_1\left(\tau,\lambda\right)=\frac{\tau\lambda}{2\left(\tau-\lambda\right)}+\frac{\tau^2\lambda^2}{2\left(\tau-\lambda\right)^2}\left[f\left(\tau\right)-f\left(\lambda\right)\right]+\frac{\tau^2\lambda}{\left(\tau-\lambda\right)^2}\left[g\left(\tau\right)-g\left(\lambda\right)\right] \\[0.1cm]
&I_2\left(\tau,\lambda\right)=-\frac{\tau\lambda}{2\left(\tau-\lambda\right)}\left[f\left(\tau\right)-f\left(\lambda\right)\right] \; , 
\end{split}
\end{equation}
where $f(\tau)$ is defined in Eq.~(\ref{eq:ftau}) and  $g(\tau)$ reads
\begin{equation}
g\left(\tau\right)=\left\lbrace \begin{array}{ll}
\sqrt{\tau-1}\, \arcsin \displaystyle\frac{1}{\sqrt{\tau}} & \tau\geq1 \\[0.5cm]
\displaystyle \frac{\sqrt{1-\tau}}{2}\left[\ln\frac{1+\sqrt{1-\tau}}{1-\sqrt{1-\tau}}-i\pi\right] & \tau<1 \, .\end{array} \right. 
\end{equation}
The QCD radiative corrections~\cite{nloZga} are small and thus
have been neglected, while the NLO EW corrections are unknown.
Because of the smallness of the QCD corrections, there is no relevant issue arising from the 
intrinsic uncertainty due to the unknown higher-order corrections, 
so that the choice of the scheme in which the quark masses
  are calculated does not play any role. In {\tt eHDECAY} we use the
  pole masses for the quarks.
Finally, Eq.~(\ref{eq:Zgammachiral}) neglects terms further suppressed by $O(m_h^2/M^2)$,
which are of higher-order in the derivative expansion. The decay width
for the MCHM4 and MCHM5 is obtained by replacing $c_\psi$ and $c_W$ with the
coupling values of column 3 and 4 of Table~\ref{tab:coupvalues} and
setting $c_{Z\gamma} \equiv 0$. 

In the SILH parametrization the decay width is computed by  {\tt
  eHDECAY} according to the formula 
\begin{equation}
\label{eq:ZgaSILH}
\begin{split}
 \Gamma (Z\gamma)\big|_{SILH} =  & \, \frac{G_F^2\alpha_{em}m_W^2m_h^3}{64\pi^4} \left(1-\frac{m_Z^2}{m_h^2}\right)^3 \\
& \times \Bigg\{  
\left|A^{SM} (Z\gamma)\right|^2 + 2 \, \text{Re} \!\left( A^{SM*} (Z\gamma)  \, \Delta A(Z\gamma) \right) \\ 
& \hspace{0.8cm} + 2 \, \text{Re} \!\left[ -\frac{4\pi
    \tan\theta_W}{\sqrt{\alpha_{em} \alpha_2}} (\bar{c}_{HB} -\bar{c}_{HW} - 8\bar{c}_\gamma \sin^2 \theta_W) 
    \, A^{SM*} (Z\gamma) \right] \Bigg\} \, , 
\end{split}
\end{equation}
where we have defined the LO SM amplitude
\begin{equation}
A^{SM} (Z\gamma) = \sum_\psi \frac{N_c Q_\psi\hat
     v_\psi}{\cos\theta_W}\, A_{1/2}^{Z\gamma}\left(\tau_\psi,\lambda_\psi\right)
    +A_1^{Z\gamma}\left(\tau_W,\lambda_W\right)
\end{equation}
and the deviation from the SM amplitude
\begin{equation}
\Delta A (Z\gamma) = - \sum_\psi \left( \frac{\bar{c}_H}{2} +
  \bar{c}_{\psi} 
\right) \frac{N_c Q_\psi\hat  v_\psi}{\cos\theta_W}\, A_{1/2}^{Z\gamma}\left(\tau_\psi,\lambda_\psi\right) 
    - \left( \frac{\bar{c}_H}{2} -2\bar{c}_W \right) \, A_1^{Z\gamma}\left(\tau_W,\lambda_W\right) \:. 
\end{equation}
Equation~(\ref{eq:ZgaSILH}) includes corrections of $O(v^2/f^2)$ and $O(m_h^2/M^2)$. The EW corrections
are unknown, and small QCD radiative corrections have been neglected.

%%%%%%%%%%%%%%%%%%%%%%%%%%%%%%%%%%%%%%%%%%%%%%%%%%%%%%%%
\subsection{Decays into $WW$ and $ZZ$ boson pairs}
%%%%%%%%%%%%%%%%%%%%%%%%%%%%%%%%%%%%%%%%%%%%%%%%%%%%%%%%
The Higgs boson decay into a pair of massive vector bosons is
important not only above the  threshold, but also below. For example,
in the SM with $m_h=125$\,GeV the branching ratio of $h\rightarrow WW$
is about 20\%. In {\tt HDECAY} various options are present to compute
the partial decay widths  
with on-shell or off-shell bosons, controlled by the ON-SH-WZ input
parameter. In {\tt eHDECAY} we have implemented the case ON-SH-WZ=0,
which includes the double off-shell decays $h\rightarrow
W^*W^*,Z^*Z^*$. For this case, which is obviously the most complete as
it takes into account both on-shell and off-shell contributions, the
partial decay width $h\rightarrow V^*V^*$ ($V=W,Z$) can be written in
the following compact form~\cite{Cahn:1988ru}:  
\begin{equation}
 \Gamma (V^*V^*)=\frac{1}{\pi^2}\int_0^{m_h^2} \!\frac{dQ_1^2\; m_V\Gamma_V}{\left(Q_1^2-m_V^2\right)^2+m_V^2\Gamma_V^2}
 \, \int_0^{\left(m_h-Q_1\right)^2} \!\!\frac{dQ_2^2 \; m_V\Gamma_V}{\left(Q_2^2-m_V^2\right)^2+m_V^2\Gamma_V^2}\
 \Gamma (VV)\, ,
\end{equation}
where $Q_1^2$, $Q_2^2$ are the squared invariant masses of the virtual
gauge bosons and $m_V$ and $\Gamma_V$ their masses and total decay
widths.  In the  parametrization of Eq.~(\ref{eq:chiralL}), by defining
\begin{equation}
\label{eq:couppar}
a_{VV} = c_{VV} \, \frac{m_h^2}{m_V^2} \, , \qquad
a_{V\partial V} = \frac{c_{V\partial V}}{2} \,  \frac{m_h^2}{m_V^2} \, ,
\end{equation}
the squared matrix element $\Gamma (VV)$ reads
\begin{equation}
\label{eq:expanded}
\begin{split}
\Gamma (VV)\big|_{NL}  = \; \Gamma^{SM} (VV) \times \Bigg\{
& c_V^2 - 2 c_V \left[ \frac{a_{VV} }{2} \left( 1-
      \frac{Q_1^2 + Q_2^2}{m_h^2} \right) + a_{V\partial V} \frac{Q_1^2 +
    Q_2^2}{m_h^2} \right]  \\
& + c_V a_{VV} \,  \frac{\lambda\left(Q_1^2,Q_2^2,m_h^2\right) \;
  (1-(Q_1^2+Q_2^2)/m_h^2)}{\lambda\left(Q_1^2,Q_2^2,m_h^2\right)+12 \, Q_1^2Q_2^2/m_h^4} \Bigg\} \; , 
\end{split}
\end{equation}
with~\cite{Cahn:1988ru}
\begin{equation}
\Gamma^{SM} (VV) =\frac{\delta_VG_Fm_h^3}{16\sqrt{2}\pi}
 \sqrt{\lambda\left(Q_1^2,Q_2^2,m_h^2\right)} \left(\lambda\left(Q_1^2,Q_2^2,m_h^2\right)
    +\frac{12Q_1^2Q_2^2}{m_h^4}\right) \;,
\end{equation}
where $\delta_V=2(1)$ for $V=W(Z)$ and $\lambda(x,y,z)\equiv
(1-x/z-y/z)^2-4xy/z^2$. The second and third term in
Eq.~(\ref{eq:expanded}) represent the  interference between the
tree-level contribution 
and the one from the derivative operators. They are of order
$O(m_h^2/M^2)$, hence next-to-leading in the chiral expansion compared
to the tree-level contribution; we have consistently neglected terms
quadratic in $a_{VV}$ and $a_{V\partial V}$, since they are of
$O(m_h^4/M^4)$, which is beyond the accuracy of the effective
Lagrangian~(\ref{eq:chiralL}). Setting $a_{VV}=a_{V\partial V}=0 $ and
$c_V= \sqrt{1-\xi}$ we obtain the decay formula for the MCHM4 and
MCHM5.
%% , which includes terms of $O(v^2/f^2$). 

In the SILH parametrization the squared matrix element $\Gamma (VV)$
implemented in {\tt eHDECAY} reads
\begin{equation}
\label{eq:VVSILH}
\Gamma (VV)\big|_{SILH}  = \Gamma^{SILH} (VV) + \Gamma^{SM} (VV) 
\frac{2}{|A_0^{SM}|^2} \, \mbox{Re} \left( A_0^{*SM} A^{SM}_{ew} \right) \;,
\end{equation}
where $A_0^{SM}$ denotes the SM LO amplitude and $A^{SM}_{ew}$ is the SM
amplitude which comprises the NLO EW corrections~\cite{nloewVV} (the
same remark as in footnote~\ref{footnote:EW} applies). Furthermore, 
\begin{equation}
\begin{split}
\Gamma^{SILH} (VV) =  \; \Gamma^{SM} (VV) \times \Bigg\{  
& 1-\bar{c}_H -2 \bar{c}_T \delta_{VZ} -2 \left[ \frac{\bar{a}_{VV} }{2} \left( 1-
      \frac{Q_1^2 + Q_2^2}{m_h^2} \right) + \bar{a}_{V\partial V} \frac{Q_1^2 +
    Q_2^2}{m_h^2} \right]  \\[0.1cm]
& + \bar{a}_{VV} \, \frac{\lambda\left(Q_1^2,Q_2^2,m_h^2\right) \,
  (1-(Q_1^2+Q_2^2)/m_h^2)}{\lambda\left(Q_1^2,Q_2^2,m_h^2\right)+12\, Q_1^2Q_2^2/m_h^4} \Bigg\} \, ,
\end{split}
\end{equation}
with $\delta_{VZ}=0 (1)$ for $V=W(Z)$ and where we have defined, 
\begin{align}
\begin{split}
\bar{a}_{WW} & = -2 \, \frac{m_h^2}{m_W^2}\, \bar c_{HW} \\[0.1cm]
\bar{a}_{W\partial W} &= -2 \, \frac{m_h^2}{2 m_W^2}  \left( \bar c_W + \bar c_{HW} \right) 
\end{split}
\\[0.5cm]
\begin{split}
\bar{a}_{ZZ} & = -2 \, \frac{m_h^2}{m_Z^2}  \left(\bar c_{HW} + \bar c_{HB} \tan^2\!\theta_W - 4 \bar c_\gamma \tan^2\!\theta_W \sin^2\!\theta_W\right) \\[0.1cm]
\bar{a}_{Z\partial Z} &= -2 \, \frac{m_h^2}{2 m_Z^2}  \left(\bar c_W + \bar c_{HW}  + (\bar c_B + \bar c_{HB}) \tan^2\!\theta_W\right) 
\end{split}
\end{align}
The decay width~(\ref{eq:VVSILH}) includes terms of $O (v^2/f^2)$,
$O(m_h^2/M^2)$ and $O (\alpha_2/4\pi)$, while it 
neglects contributions of $O(v^4/f^4)$ and $O[(\alpha_2/4\pi)^2]$. The
corrections of $O[(\alpha_2/4\pi)(v^2/f^2)]$ are only partly included
through the terms proportional to $\bar c_{HW}$ and~$\bar c_{HB}$,
{\it cf.}~\cite{Contino:2013kra}.

%%%%%%%%%%%%%%%%%%%%%%%%%%%%%%%%%%%%%%%%%%%%%%%%%%%%%%%%
\section{Numerical formulas for the decay rates in the SILH Lagrangian}
\label{sec:numerical}
%%%%%%%%%%%%%%%%%%%%%%%%%%%%%%%%%%%%%%%%%%%%%%%%%%%%%%%%
We display here numerically approximated formulas of the Higgs decay
rates valid at linear order in the effective coefficients $\bar c_i$
of the SILH Lagrangian~(\ref{eq:silh})  for $m_h=125\,$GeV. 
All the ratios $\Gamma/\Gamma_{SM}$ have been computed by switching
off the EW corrections, since their effect on the numerical
prefactors appearing in front of the coefficients $\bar c_i$ is of
order $(v^2/f^2)(\alpha_2/4\pi)$ and thus beyond the accuracy of the
formulas implemented in {\tt eHDECAY}. Conversely, we have fully
included the QCD corrections, as they multiply both the SM and the NP terms.
The numerical results are thus the following:
\begin{align}
\frac{\Gamma(\bar\psi\psi)}{\Gamma(\bar\psi\psi)_{SM}} \simeq & \, 1 -
\bar c_H - 2\, \bar c_\psi \, , \quad \mbox{for } \psi = \mbox{leptons,
top-quark} \;.
\end{align}
The QCD corrections to the decays decays into charm, strange or bottom
quark pairs involve two-loop diagrams with top quarks loops that are
rescaled differently~\cite{Chetyrkin:1995pd}. Taking
this into account, we have the numerical 
results
\begin{align}
\frac{\Gamma(\bar{c} c)}{\Gamma(\bar{c} c)_{SM}} \simeq & \, 1 -
\bar c_H - 1.985\, \bar c_c  - 0.015 \, \bar{c}_t \, , \\[0.5cm]
\frac{\Gamma(\bar{s} s)}{\Gamma(\bar{s} s)_{SM}} \simeq & \, 1 -
\bar c_H - 1.971\, \bar c_s - 0.029 \, \bar{c}_t \, , \\[0.5cm]
\frac{\Gamma(\bar{b} b)}{\Gamma(\bar{b} b)_{SM}} \simeq & \, 1 -
\bar c_H - 1.992\, \bar c_b - 0.0085 \, \bar{c}_t \, .
\end{align}
Furthermore, 
\begin{align}
\label{eq:numWW}
\frac{\Gamma(h\to W^{(*)}W^{*})}{\Gamma(h\to W^{(*)}W^{*})_{SM}}
\simeq & \, 1 - \bar c_H + 2.2 \, \bar c_W + 3.7 \, \bar c_{HW} \, ,
\\[0.5cm] 
\label{eq:numZZ}
\begin{split}
\frac{\Gamma(h\to Z^{(*)}Z^{*})}{\Gamma(h\to Z^{(*)}Z^{*})_{SM}}  \simeq 
& \, 1 - \bar c_H -2 \bar c_T + 2.0 \, \left( \bar c_W + \tan^2\!\theta_W \, \bar
  c_B\right)  \\
&+ 3.0 \, \left(\bar c_{HW} + \tan^2\!\theta_W \, \bar c_{HB}\right)
- 0.26 \, \bar c_\gamma\, ,  
\end{split}
\\[0.5cm]
\begin{split}
\frac{\Gamma(h\to Z\gamma)}{\Gamma(h\to Z\gamma)_{SM}}  \simeq 
& \, 1 - \bar c_H + 0.12 \, \bar c_t - 5 \cdot 10^{-4} \, \bar c_c- 0.003
\, \bar c_b - 9 \cdot 10^{-5} \, \bar c_\tau \\ 
& + 4.2 \, \bar c_W + 0.19 \left( \bar c_{HW} - \bar c_{HB} + 8\, \bar
  c_\gamma \sin^2\!\theta_W \right)  
\frac{4\pi}{\sqrt{\alpha_2 \alpha_{em}}}\, , 
\end{split} \\[0.5cm]
\begin{split}
\frac{\Gamma(h\to \gamma\gamma)}{\Gamma(h\to \gamma\gamma)_{SM}}
\simeq 
& \, 1 - \bar c_H + 0.54 \, \bar c_t - 0.003 \, \bar c_c - 0.007 \,
\bar c_b - 0.007 \, \bar c_\tau   \\
& + 5.04 \, \bar c_W - 0.54 \, \bar c_\gamma \, \frac{4\pi}{\alpha_{em}} \, , 
\end{split} \\[0.5cm]
\frac{\Gamma(h\to gg)}{\Gamma(h\to gg)_{SM}}  \simeq 
& \, 1 - \bar c_H - 2.12 \, \bar c_t + 0.024 \, \bar c_c + 0.1 \, \bar c_b + 22.2 \, \bar c_g \, \frac{4\pi}{\alpha_{2}} \, .
\end{align}

%%%%%%%%%%%%%%%%%%%%%%%%%%%%%%%%%%%%%%%%%%%%%%%%%%%%%%
\section{How to run {\tt eHDECAY}: Input/Output Files}
\label{sec:IO}
 %%%%%%%%%%%%%%%%%%%%%%%%%%%%%%%%%%%%%%%%%%%%%%%%%%%%%%
The program {\tt eHDECAY} is self-contained, like the original code
{\tt HDECAY} on which it is based. All the new features related to the
Lagrangian parametrizations proposed in this paper are encoded in the main
source file, ehdecay.f, while other linked routines are taken over
from the original version. 
Of course {\tt eHDECAY}, besides calculating Higgs branching ratios and
decay widths according to the non-linear, SILH or MCHM4/5 Lagrangians, also
calculates the SM and MSSM ones, exactly as {\tt HDECAY}~5.10 does. The
choice can be done through the flags HIGGS and COUPVAR set in the
input file. The input file for {\tt eHDECAY} has been called ehdecay.in and is
based on the file hdecay.in of the official version 5.10, supplemented
by further input values. Thus, with the flag LAGPARAM the user can
choose between the 
general SILH parametrization Eq.~(\ref{eq:silh}), the model-specific
parametrizations MCHM4 and MCHM5 and the general non-linear Lagrangian
parametrization Eq.~(\ref{eq:chiralL}). Furthermore, the various
related couplings can be set.   
The input values are explained in the following: \\

\noindent
\underline{COUPVAR, HIGGS}: If HIGGS=0 and COUPVAR=1, then the Higgs
decay widths and branching ratios are calculated within the
parametrization chosen by: \\ 

\noindent
\underline{LAGPARAM}:

0: Non-linear Lagrangian parametrization Eq.~(\ref{eq:chiralL})

1: SILH parametrization Eq.~(\ref{eq:silh})

2: MCHM4/5 parametrization ({\it cf.}~Table~\ref{tab:coupvalues}) \\

\noindent
 \underline{IELW}: Turn off (0) or on (1) the electroweak corrections
 for the SILH parametrization.~\footnote{Note, that this
  parameter IELW has nothing to do with the parameter ELWK in the
  input file of {\tt HDECAY}, where the meaning of this flag is different.}\\

\noindent For the non-linear Lagrangian the following parameters have to
be set for the couplings of the various vertices:\footnote{We explain them here all, although  they are in part already present in the input file for {\tt HDECAY} 5.10.}
 \\

\begin{tabular}{lll}
\underline{CV}: $hVV$ vertex, (V=W, Z) &
\underline{Ctau}: $h\tau\tau$ vertex &
\underline{Cmu}: $h\mu\mu$ vertex \\[0.1cm]
\underline{Ct}: $ht\bar t$ vertex &
\underline{Cb}: $hb\bar b$ vertex &
\underline{Cc}: $hc\bar c$ vertex \\[0.1cm] 
\underline{Cs}: $hs\bar s$ vertex &
\underline{Cgaga}: coupling $c_{\gamma\gamma}$ &
\underline{Cgg}: coupling $c_{gg}$\\[0.1cm]
\underline{CZga}: coupling $c_{Z\gamma}$ &
\underline{CWW}: coupling $c_{WW}$ &
\underline{CZZ}: coupling $c_{ZZ}$ \\[0.1cm]
\underline{CWdW}: coupling $c_{W\partial W}$ &
\underline{CZdZ}: coupling $c_{Z\partial Z}$
\end{tabular}

\vspace*{0.8cm}
\noindent
In case of the SILH parametrization the input values to be set in
order to calculate the various couplings are:\\

\begin{tabular}{lllll}
\underline{CHbar}: $\bar{c}_H$ &
\underline{CTbar}: $\bar{c}_T$ &
\underline{Ctaubar}: $\bar{c}_\tau$&
\underline{Cmubar}: $\bar{c}_\mu$ &
\underline{Ctbar}: $\bar{c}_t$  \\[0.1cm]
\underline{Cbbar}: $\bar{c}_b$ &
\underline{Ccbar}: $\bar{c}_c$ &
\underline{Csbar}: $\bar{c}_s$ &
\underline{CWbar}: $\bar{c}_W$ & 
\underline{CBbar}: $\bar{c}_B$ \\[0.1cm]
\underline{CHWbar}: $\bar{c}_{HW}$ &
\underline{CHBbar}: $\bar{c}_{HB}$ &
 \underline{Cgambar}: $\bar{c}_{\gamma}$ &
\underline{Cgbar}: $\bar{c}_{g}$ 
\end{tabular}
 
\vspace*{0.8cm}
\noindent
In the MCHM4/5 parametrization we have the input values: \\

\noindent
\underline{FERMREPR}: 

1: MCHM4

2: MCHM5 \\

\noindent
\underline{XI}: the value for $\xi$ \\

\noindent
For example: \\

\noindent\texttt{COUPVAR  = 1 \\
HIGGS    = 0 \\
$\vdots$ \\
************** LAGRANGIAN 0 - chiral  1 - SILH  2 - MCHM4/5 **************\\
LAGPARAM = 0\\
**** Turn off (0) or on (1) the elw corrections for LAGPARAM = 1 or 2 ****\\
IELW     = 1\\
******************* VARIATION OF HIGGS COUPLINGS*************************\\
CW        = 1.D0 \\
CZ         = 1.D0  \\
Ctau      = 0.95D0 \\
Cmu      = 0.95D0 \\
Ct          = 0.95D0 \\
Cb         = 0.95D0 \\
Cc          = 0.95D0 \\
Cs          = 0.95D0 \\
Cgaga    = 0.005D0 \\
Cgg        = 0.001D0 \\
CZga      = 0.D0 \\
CWW      = 0.D0 \\
CZZ        = 0.D0 \\
CWdW     = 0.D0 \\
CZdZ      = 0.D0 \\
$\vdots$ \\}
computes the branching ratios for $c_V=1$, 
$c_\psi=0.95$ ($\psi=t,b,c,s,\tau,\mu$), $c_{\gamma\gamma}=0.005$, $c_{gg}=0.001$ and
$c_{Z\gamma}=c_{WW}=c_{ZZ}=c_{W\partial W}=c_{Z\partial Z}=0$ in the
general parametrization Eq.~(\ref{eq:chiralL}). The output is written
into the files br.eff1 and br.eff2, where the Higgs mass, branching
ratios and total width are reported. For the previous example, at
$m_h=125$\,GeV and for all the other parameters set at their standard values, the output reads

\begin{verbatim}
   MHSM        BB       TAU TAU     MU MU         SS         CC         TT 
_______________________________________________________________________________

 125.000     0.5895     0.5654E-01 0.2002E-03 0.2161E-03 0.2569E-01   0.000
\end{verbatim}
\newpage
\begin{verbatim}
   MHSM          GG     GAM GAM     Z GAM         WW         ZZ       WIDTH
_______________________________________________________________________________

 125.000     0.9611E-01 0.1932E-03 0.1526E-02   0.2045   0.2554E-01 0.4129E-02
\end{verbatim}
All the input parameters of the corresponding run are printed out in
the file br.input. Otherwise, setting COUPVAR=0, the program produces
the usual output files with SM or MSSM results according to the {\tt
  HDECAY}~5.10 version.

%%%%%%%%%%%%%%%%%%%%%%%%%%%%%%%%%%%%%%%%%%%%%%%%%%%%%%
\section{Conclusions \label{sec:concl}}
%%%%%%%%%%%%%%%%%%%%%%%%%%%%%%%%%%%%%%%%%%%%%%%%%%%%%%

We have described the Fortran code {\tt eHDECAY}, which calculates the partial
widths and the branching fractions of the decays of the  Higgs boson
in the Standard Model and its extension by the dimension-6 operators
of the SILH Lagrangian~(\ref{eq:silh}). The program also implements
the more general non-linear effective Lagrangian~(\ref{eq:chiralL}),
which does not rely on assuming the Higgs boson to be part of an
$SU(2)_L$ doublet. In the SM, all decay modes are included as in the
original version of {\tt HDECAY}. The corrections due to the effective
operators have been included consistently with the multiple
perturbative expansion in the number of derivatives, fields and SM
couplings. The  level of approximation of  the formulas implemented in
{\tt eHDECAY} has been discussed in detail for each decay final
state. The QCD corrections to the hadronic decays as well as the
possibility of virtual intermediate states have been incorporated
according to the present state of the art. The QCD corrections are
assumed to factorize also in the presence of higher-dimension
operators, so that they are included in factorized form in all
extensions of the SM. For the SILH case we have added the
electroweak corrections to the SM part only and left the dimension-6
contributions at LO in the context of electroweak corrections, since
deviations from the SM case are assumed to be small. In the case of
the non-linear Lagrangian however, deviations can be large so that the
non-factorizing electroweak corrections to the SM part are subleading
and thus have not been taken into account for consistency. 

The program is fast and can be used easily. The basic SM and
SILH/non-linear input parameters can be chosen from an input file.
Examples of output files for the decay branching ratios have been given.

Since electroweak corrections involving the novel operators have not
been calculated yet, the treatment of this type of corrections is not
complete. During the coming years one may expect that these electroweak
corrections will be determined so that the existing code {\tt eHDECAY} can be
extended to incorporate them.
For the moment the present version of {\tt eHDECAY} provides the state-of-the art for
the partial Higgs decay widths and branching ratios in extensions of the
SM by a SILH or a non-linear effective Lagrangian.

%%%%%%%%%%%%%%%%%%%%%%%%%%%%%%%%%%%%%%%%%%%%%%%%%%%%%%
\section*{Acknowledgments}
%%%%%%%%%%%%%%%%%%%%%%%%%%%%%%%%%%%%%%%%%%%%%%%%%%%%%%

We thank Belen Gavela, Concepcion Gonzalez-Garcia and Alex Pomarol, for useful
discussions. This research has been partly supported by the European
Commission under the ERC Advanced Grant 226371 MassTeV and the
contract PITN-GA-2009-237920 UNILHC. C.G. is  supported by the Spanish
Ministry MICNN under contract FPA2010-17747. 
The work of R.C. was partly supported by the ERC Advanced Grant No. 267985 
\textit{Electroweak Symmetry Breaking, Flavour and Dark Matter: One
  Solution for Three Mysteries (DaMeSyFla)}.
The work of M.G. has been partly supported by MIUR under contract
2001023713$\_$006. 
M.M. is supported by the DFG SFB/TR9 Computational Particle Physics.

%%%%%%%%%%%%%%%%%%%%%%%%%%%%%%%%%%%%%%%%%%%%%%%%%%%%%%
%%%%%%%%%%%%%%%%%%%%%%%%%%%%%%%%%%%%%%%%%%%%%%%%%%%%%%
%% BIBLIOGRAPHY
%%%%%%%%%%%%%%%%%%%%%%%%%%%%%%%%%%%%%%%%%%%%%%%%%%%%%%
%%%%%%%%%%%%%%%%%%%%%%%%%%%%%%%%%%%%%%%%%%%%%%%%%%%%%%

\end{document}